# A Digital Twin to overcome long-time challenges in Photovoltaics

Larry Lüer[1], Marius Peters[2], Ana Sunčana Smith[3,4], Eva Dorschky[5], Bjoern M. Eskofier[5], Frauke Liers[6], Jörg Franke[7], Martin Sjarov[7], Matthias Brossog[7], Dirk Guldi[8], Andreas Maier[9] and Christoph Brabec[1,2]

[1] Institute of Materials for Electronics and Energy Technology (i-MEET), Friedrich-Alexander-Universität Erlangen-Nürnberg, Martensstrasse 7, 91058 Erlangen, Germany

[2] High Throughput Methods in Photovoltaics, Forschungszentrum Jülich GmbH, Helmholtz Institute Erlangen-Nürnberg for Renewable Energy (HI ERN), Immerwahrstraße 2, 91058 Erlangen, Germany

[3] PULS Group, Department of Physics, Friedrich-Alexander-Universität Erlangen-Nürnberg, IZNF, Erlangen, Germany.

[4]Division of Physical Chemistry, Ruđer Bošković Institute, Bijenička cesta 54, Zagreb, Croatia.

[5] Machine Learning and Data Analytics Lab, Friedrich-Alexander-Universität Erlangen-Nürnberg, Carl-Thiersch-Straße 2b, 91052 Erlangen, Bayern, Germany

[6] Department of Data Science (DDS), Friedrich-Alexander-Universität Erlangen-Nürnberg, Cauerstr. 11, 91058 Erlangen, Germany

[7] Institute for Factory Automation and Production Systems (FAPS), Friedrich-Alexander-Universität Erlangen-Nürnberg, Egerlandstr. 7, 91058 Erlangen, Germany

[8] Department of Chemistry and Pharmacy, Egerlandstr. 3, Friedrich-Alexander-Universität Erlangen-Nürnberg 91058 Erlangen, Germany

[9] Informatics Department, Friedrich-Alexander-Universität Erlangen-Nürnberg, Martenstr. 3, 91058 Erlangen, Germany

## Abstract

The recent successes of emerging photovoltaics (PV) such as organic and perovskite solar cells are largely driven by innovations in material science. However, closing the gap to commercialization still requires significant innovation to match contradicting requirements such as performance, longevity and recyclability. The rate of innovation, as of today, is limited by a lack of design principles linking chemical motifs to functional microscopic structures, and by an incapacity to experimentally access microscopic structures from investigating macroscopic device properties. Both limitations in turn are caused by an individualist approach to learning, not being able to produce consistent datasets large enough to find patterns and physical laws leading us to breakthrough innovations. In this work, we envision a layout of a Digital Twin for PV materials aimed at removing both limitations.

The layout combines machine learning approaches, as performed in materials acceleration platforms (MAPs), with mathematical models derived from the underlying physics and digital twin concepts from the engineering world. This layout will allow using high-throughput (HT) experimentation in MAPs to improve the parametrization of quantum chemical and solid-state models. In turn, the improved and generalized models can be used to obtain the crucial structural parameters from HT data. HT experimentation will thus yield a detailed understanding of generally valid structure-property relationships, enabling inverse molecular design, that is, predicting the optimal chemical structure and process conditions to build PV devices satisfying a multitude of requirements at the same time. After motivating our proposed layout of the digital twin with causal relationships in material science, we discuss the current state of the enabling technologies such as model-based inference of hidden parameters, already being able to yield insight from HT data today. However, we identify the multiscale nature of PV materials and the needed volume and diversity of data as main challenges. This requires the development of novel methods for scale bridging fast surrogates based on physical models, which in turn requires novel optimization methods being able to handle continuous optimization and discrete decisions in large instances under uncertainty. We mention promising approaches to address these challenges.

## Introduction

Emerging photovoltaics (PV) technologies such as organic and perovskite solar cells (OSC and PSC, respectively) have seen a rapid increase of efficiencies over the last decade, which was largely driven by material science. The power conversion efficiency (PCE) of single-junction organic solar cells has been pushed beyond 19% [1-3]. Large area modules have also demonstrated PCEs of over 12% [4-6]. These values are however not yet sufficient for volume markets. Moreover, important issues regarding operational stability [7-13] and up-scaling [14,15] still prevail. Although there is progress [16-18], it is too slow to attract venture capital to bridge the gap between lab scale and industry scale production, given the presence of a very competitive established technology today (Silicon PV) and the promise of quasi-infinite energy availability at the horizon (nuclear fusion).

These points can be quantitatively demonstrated by the so-called learning rate (LR), which tracks €/Wp as a function of accumulated production volume. We know that silicon (Si) and cadmium telluride (CdTe) PV technologies have a LR of 20% – 25% (see Figure 1a) [19]. Therefore, any emerging PV technology, depending on its starting point which is usually reported at the MW scale level, must have a faster LR to catch up with the Si and CdTe PV technologies in terms of cost competitiveness. As one example to underline the importance of acceleration, we plotted the cost evolution for a presumably carbon-based PV technology, assuming a doubling of the LR. Figure 1a shows that a learning rate of 50 % would reduce the required cumulative production volume for becoming a mass market-ready PV technology by about a factor of 1000, that is, from about 100 GW to only 100 MW. This means that the first serious production line would be cost competitive from the beginning.

The learning rate is increased by scaling effects coming from industrial volume production, but also by innovation and discovery coming from R&D. To speed up discovery in material science, so-called Materials Acceleration Platforms (MAP) [20] are gaining momentum. In MAP, design-of experiment is performed by strategically dividing the quasi-infinite experimental space in hierarchically connected subspaces, in which optimization and knowledge generation is performed by screening or active learning (see Figure 1b). The hierarchy is given by the scales of the underlying physical principles: on the (supra)molecular scale, the connection between chemical structure and molecular properties such as energies, vibronic coupling ("Energy gap law", Marcus-Levich-Jortner transfer rates), and inter-molecular energy level alignment (excitons, charge transport) can be predicted by quantum chemistry more quickly than measured experimentally. One can therefore screen large numbers of candidate molecules virtually [21], and, by assessing in parallel the resulting molecular quantities experimentally for only a small fraction, will assist to improve the calculated results until a small number of optimal candidate molecules are found. These molecules pass to the next stage, in which the physics of the mesoscopic scale is optimized. Currently, this stage is largely experimental because prediction of microstructure formation and phase separation [22] as well as simulation of photoexcitation dynamics therein [23, 24] is slower than actually performing the experiment. Using fast proxy experiments for device performance [25] or stability [26], large numbers of single layers can be screened in shortest time, resulting in recipes for optimal layer formulations. These can pass to the final stage, in which the physics of the macroscopic scale is optimized. In this stage, the interplay of the layers is optimized to maximize the final performance indicators.

Although these MAPs have shown progress against the state of the art, breakthrough innovations have so far not been presented. Figure 1b exposes the main bottleneck to disruptive discovery in the current design of MAPs, namely the lack of insight in the mesoscopic stage. It is hard to assess structural details from high throughput capable characterization techniques. As long as the middle stage is of black box nature, communication with the adjacent stages will be lossy: neither can we encounter the nexus between molecular structure and mesoscopic structure, nor can we find fast proxy experiments for mesoscopic structure formation, which would allow us to predict the performance of the single layers in the device stack. As a consequence, sub-optimal molecules may pass the first gate because we are not able to predict their mesoscopic structure, and wrongly optimized layers will pass into device formulation because they were characterized by proxy experiments insufficiently predicting the crucial parameters for macroscopic device functionality. Although it may in principle be possible to skip the second stage using recently developed generative models for inverse molecular design, the required amount of training data will be unmanageable, as the latent space will be of much higher dimensionality for predicting complete devices than for predicting details on a mesoscopic structure level.

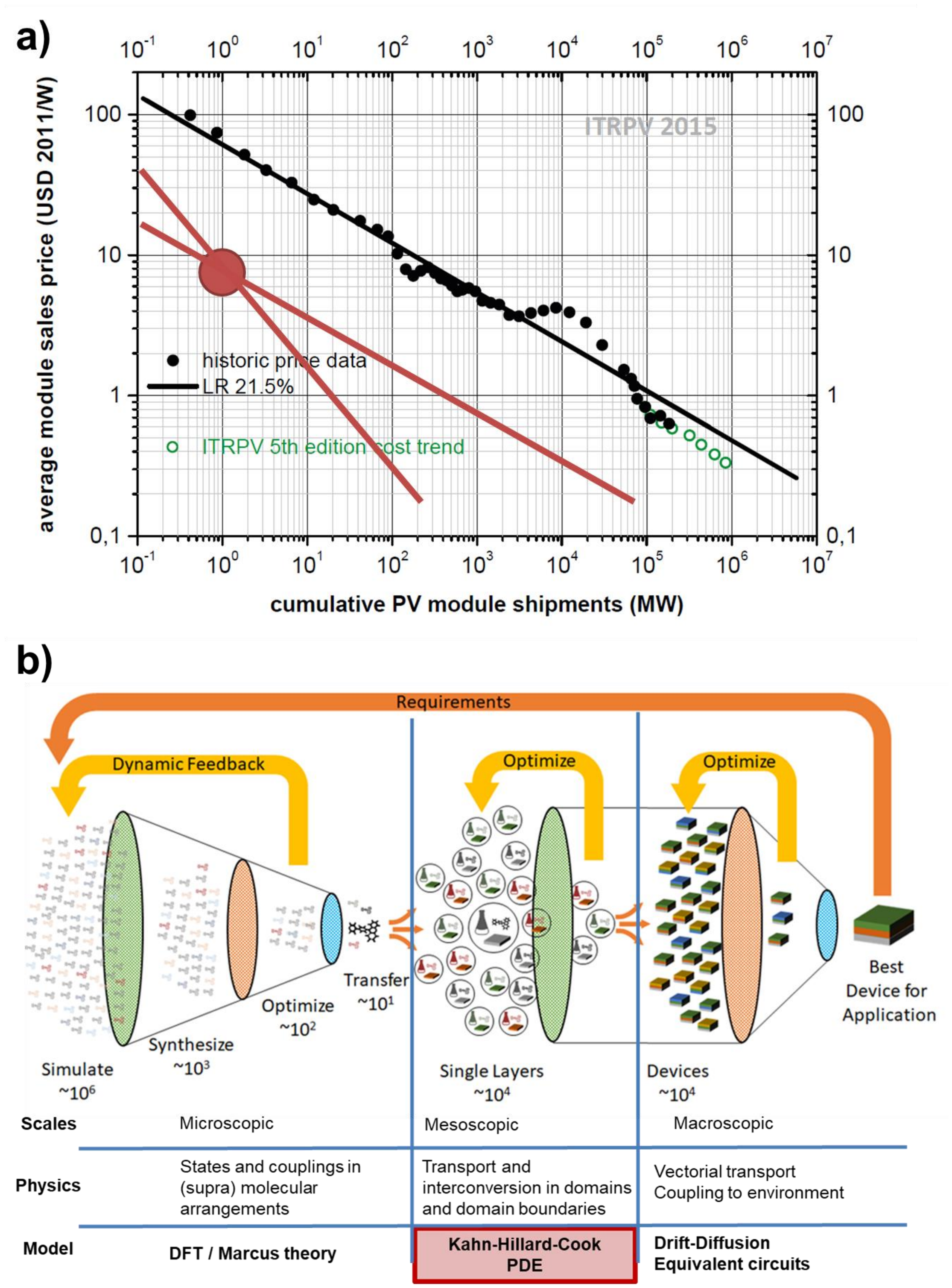

Figure 1 a) Learning rate in silicon (Si-)PV technology (green symbols & black line; exponential fit given by red line yielding 25% price reduction every time the installed capacity doubles, data from Fraunhofer ISE); the blue dashed lines shows a prospective learning rate of OPV of 50% (LR50). b) Operation principle of Material Acceleration platforms, together with the relevant length scales, physical principles and available models.

Disruptive discovery is intimately linked to the ability to attain control over principles with opposite effects on the desired targets. In disordered solids, which are the basis of emerging-PV technologies, this quest for control leads to specific challenges, spanning across all length scales, from single molecules to PV modules:

Time and energy management of electronic states. Optoelectronics relies on an interplay of different materials, and thus the respective driving forces for the desired transfer phenomena must be at their optimum values to suppress omnipresent loss processes [27]. This is particularly important in the field of multiphoton processes for PV (photon up-conversion or singlet fission [28]), where the optical absorption bandgaps, emission energies, ionization potentials and electron affinities all must match to exploit their potential to go beyond the detailed balance limit for single junction solar cells. Not only the state energies, but also their coupling must be managed, minimizing internal conversion to ground states (beyond the energy gap law [29]) but maximizing desired transfer rates. On the supramolecular scale, this requires control of intermolecular or inter-chain alignment.

Microstructure control. Maximizing optoelectronic performance in disordered semiconductors requires judicious management of nano- to microscopic arrangements, for example by creation of nanofibrils allowing swift conduction along the stack while at the same time maximizing charge generation orthogonal to it [30]. The formation of such kind of highly advantageous nanostructures can currently not be predicted from molecular structures or process conditions, often because these are metastable states. The same holds for phase composition (vertical gradients, dual phase acceptors).

Device stability. On the macroscopic level, environmental stability is key to commercial success but also to live up to the promise of sustainability – materials of different roles (electron transport, absorber, etc) may react with each other, radicals may autocatalyze degradation. Design principles for long term stability must be understood in much greater detail, given the multi-objective nature of the requirements for commercialization.

In this perspective, we argue that in order to achieve breakthrough innovations in PV material science, we need generic predictive power to address all challenges, including microstructure, so as to make contradicting principles act in concert. We present the layout of a Digital Twin in PV materials, able to accelerate the learning of microscopic structure property relationships in a combined data- and model- driven approach. High-throughput experimentation is used to improve the parametrization of simulation methods, and the improved simulations are in turn used to obtain the crucial structural parameters and drive high-throughput experiments. In this way, deeper understanding for the selection of particular molecular motifs can be acquired. This understanding can in turn lead to new optimisation designs and the establishment of new regions of interest for analysis both in modelling and the HT approaches. Hence, Collecting a vast library of molecular structures and corresponding features of mesoscopic structure, the Digital Twin will accelerate the discovery of molecular structures conducive to any of the crucial structural aspects at larger scales, compared to a purely data driven approach If any structural aspect can be individually optimized at will, we will have achieved molecular inverse design capacity, allowing to discover optimal molecules satisfying all requirements at once. Subsequently, processes for their production can be optimised.

In the following, we will first introduce into predictive models and their limits. Based on these findings we will discuss the proposed layout of the Digital Twin for PV materials, which will be followed by a detailed look at the current state of the enabling technologies.

# 1 Forward and Inverse Design

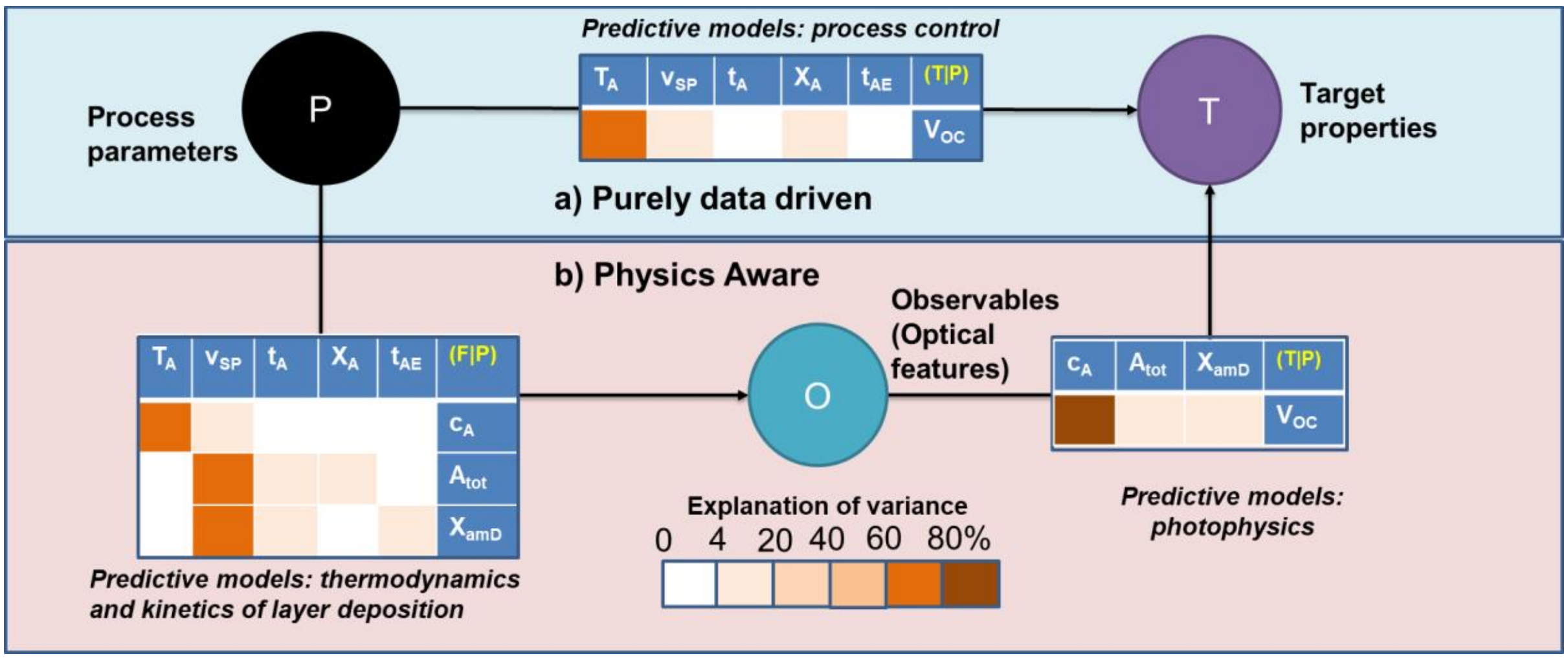

*Figure 2: Use of predictive models for the prediction of $V_{OC}$ in PM6:Y6 OPV devices. Color codes in the matrices show the explanation of variance for relevant and non-redundant predictors, obtained by mRMR-GPR, from the dataset published in [31]. Predictors are across columns, targets across rows. The predictors are explained in the text.*

In the following, we take a Bayesian perspective for the layout of a Digital Twin as a natural approach in this context and start from the establishment of a fully data-based context from which we aim at generating knowledge. We will however also aim at integrating knowledge from structural and from physical insights. Through experimentation, we seek to learn quantitative structure-property relationships (QSPR) of general validity. These can be expressed by predictive models of the form of conditional probability distributions $p(\mathbf{B}|\mathbf{A})$ for the occurrence of a certain target **B** given a vector of features (predictors) **A.** These predictive models correspond to the underlying deterministic functions **B**=f(**A**), imposed by laws of physics. Materials acceleration platforms (MAP), as discussed in the introduction, are ideally suited for this task because they allow collecting large amounts of data under controlled variation of process variables **P**, keeping variation of other (hidden) process variablesto a minimum. Using our automated platform for the processing and characterization of organic solar cells (AMANDA Line One), we have obtained predictive models for electrical performance and even operational stability for the donor-acceptor system PM6:Y6 [31]. To this end, we varied the active layer annealing temperature ($T_A$) and time ($t_A$), the acceptor molar fraction ($X_A$), the spinning speed of the spin coater ($v_{sp}$), and the annealing time of the electron transport layer ($t_{AE}$), among others. To demonstrate the construction of predictive models, we have re-examined the dataset from [31] using a feature selection technique (minimum Redundancy Maximum Relevance [32]) embedded in Gaussian Process Regression (GPR). Figure 2a shows the predictive model $p(V_{OC}|\mathbf{P})$, where the active layer annealing temperature $T_A$ has a very strong influence on the open circuit voltage $V_{OC}$, explaining

more than 80% of the variance in $V_{OC}$ of the dataset. Other parameters such as $v_{sp}$ and $X_A$ add less than 20% of additional explanation of variance, and the two annealing times $t_A$ and $t_{AE}$ do not seem to play a significant role for $V_{OC}$ in the actual dataset.

Inverse design, in a probabilistic definition, is the inverse function of a predictive model, namely $p(\mathbf{A}|\mathbf{B})$, the conditional probability of a feature given the desired value of **B** [33]. In the example of Figure 2a, the inverse design function would be $p(T_A|V_{OC})$, meaning that we are able to adjust the annealing temperature of the active layer in order to fine tune a desired open circuit voltage of the final device, for example to match the requirements of an internet of things (IoT) application. Although this is useful, it will work only for active layers composed of PM6:Y6, and only for the specific way by which AMANDA Line One processes devices.

We have shown in [31] that by adding physics to models trained from data, we can improve their generalization. Using the exciton theory and the Spano model of weak H aggregates [34], we have obtained morphology sensitive features for the active layer from modelling of UV-Vis spectra, taken before electrode deposition. The predictive model for $V_{OC}$ in Figure 2b, right, shows that $V_{OC}$ is determined of observable features **O** like the acceptor energy level $c_A$, the total absorption of acceptor molecules $A_{tot}$, and the amount of amorphous phase in the donor phase ($X_{amD}$). The predictive model $p(\mathbf{T}|\mathbf{O})$, obtained by adding models obtained from laws of physics, is more general than the purely data driven model $p(\mathbf{T}|\mathbf{P})$ from Figure 1a, because the former relates morphological features to the target property, irrespective by which processing method they have been obtained. The predictive model $p(\mathbf{T}|\mathbf{O})$, should therefore, under certain conditions, be transferrable to different processing methods, while $p(\mathbf{T}|\mathbf{P})$ is not.

In spite of the improved generalization, physics-aware predictive models as the ones in Figure 2b still do not lend themselves to breakthrough innovations. This is due to the well-known inverse problem: varying materials and process conditions, there are infinitely many possibilities to produce a certain value of $c_A$, but generally they will produce very different values of $V_{OC}$ and vice versa. The high correlation between $c_A$ and $V_{OC}$ in the present dataset stems from the fact that we varied only the process conditions but not the chemical structure of the materials in the active layer. This is shown in the ESI, Figure S1, as red arrow identifying the non-causal pathway (correlation) between $c_A$ and $V_{OC}$. In fact, one of the breakthrough innovations sought after by the OPV community is breaking the correlation between the charge transfer energy level represented by $c_A$ for hybrid interfaces and $V_{OC}$, by discovering materials with lower non-radiative voltage losses. To achieve this by inverse design, knowledge of structural details is mandatory, but is not provided by the simple predictor $c_A$ in Figure 2b.

Figure 3 shows the causal relationships in material science yielding the observed predictive models, utilizing ideas published in [35]. Process conditions **P** allow us to interact with matter in a controlled way, such that chemical structure (**C**) is formed and organized into a system **S**. The system **S** comprises the complete geometrical, energetic and dynamic structure across all scales of our asset, which can be a thin film, a single research device, or a complete array of PV modules connected to the grid. All aspects of the structure of the asset together determine the desired target properties **T**, such as the electrical performance, the longevity of the device or the purity of the recycled products. Furthermore, the structure also determines the observable response (**O**) if we interrogate the asset by spectroscopy, microscopy, or electrodynamics. Hence, if **P** is known, then both **C** and **S** can be predicted. In turn, if **S** is known, both **O** and **T** can be predicted, explaining our predictive model $p(\mathbf{T}|\mathbf{O})$ from Figure 2 as a non-causal pathway involving **S**, see bold red arrow in Figure 3. We note that these considerations are valid only in the ensemble approximation where large numbers of quantum objects yield deterministic expectation values (correspondence principle, e.g., mapping statistical mechanics onto thermodynamics in Fig. 3) and where the expectation values can be deterministically captured (ideal experiment). Both conditions hold only approximately, and the uncertainty that comes along with it must be quantified.

The problem for inverse design is the quasi-infinite dimensionality of **P**, **C**, and **S**, making the general prediction of **T** from **O** intractable. If we cannot generalize the learned models, we are bound to what we already know, which permits only small improvements but no breakthrough innovations. On the other hand, the quasi-infinite dimensionality of **P**, **C**, and **S** holds also a promise, namely the high probability that there exist a combination of **S** leading to unseen **T**, that is, to match opposing requirements, crucial to overcome the challenges mentioned in the introduction. Likewise, we expect a high probability for the existence a combination of **P** and **C** to yield said **S**.

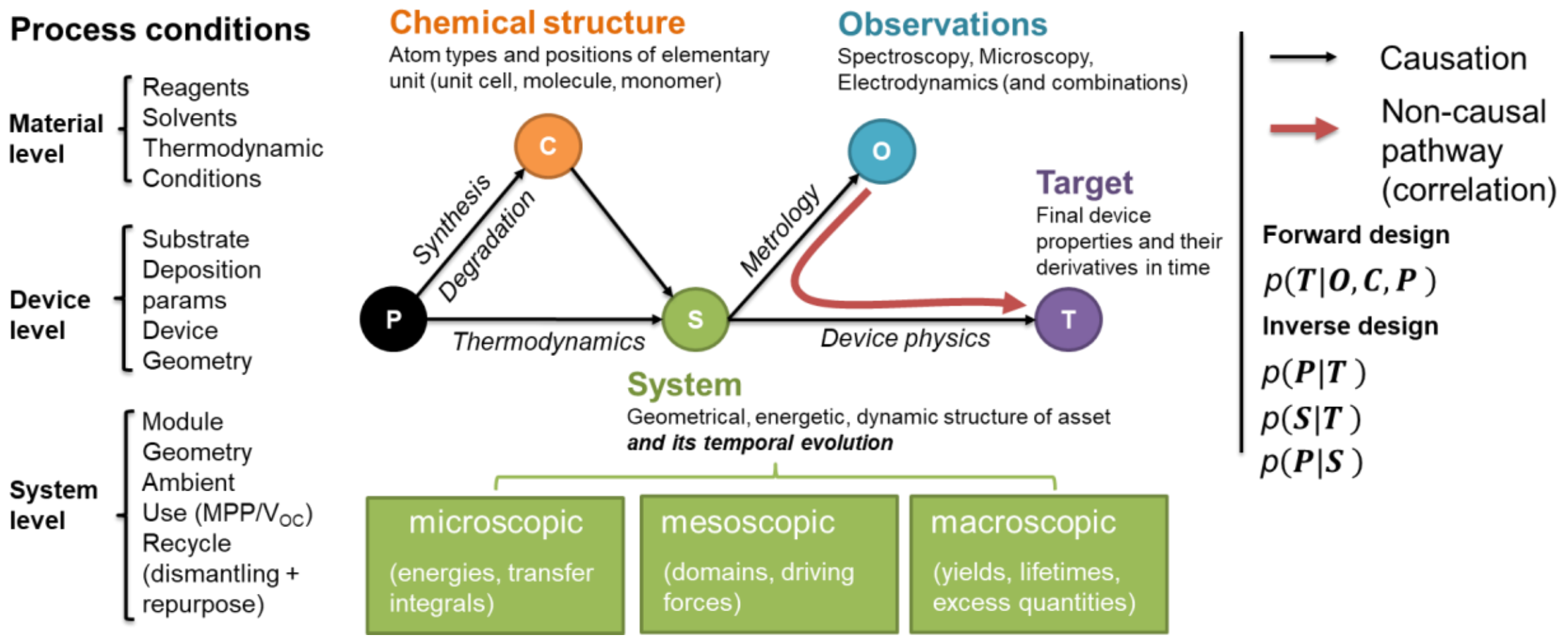

*Figure 3. Acyclic directed graph showing the causal relationships needed to achieve quantitative and generalized predictive capacity, allowing to perform molecular inverse design*

For our layout of a Digital Twin for PV materials, we propose to obtain $p(\mathbf{P},\mathbf{C}|\mathbf{T})$ in two sequential steps: on the one hand, we encounter the inverse design function $p(\mathbf{S}|\mathbf{T})$ from experimental data, where we develop fast surrogates to access microscopic system parameters. On the other hand, we develop integrated model-based and machine learning (ML)-enhanced simulations to learn the inverse design function $p(\mathbf{P},\mathbf{C}|\mathbf{S})$, that is, to predict microscopic and mesoscopic structure from chemical structure. In this way, a high dimensional problem is converted into two lower dimensional problems, which guarantees an increase of the learning rate compared to the pure data driven approach; generality is not lost, as all causal pathways connecting **P**,**C** and **T** run via **S**.

# 3 Layout of a Digital Twin for PV Materials

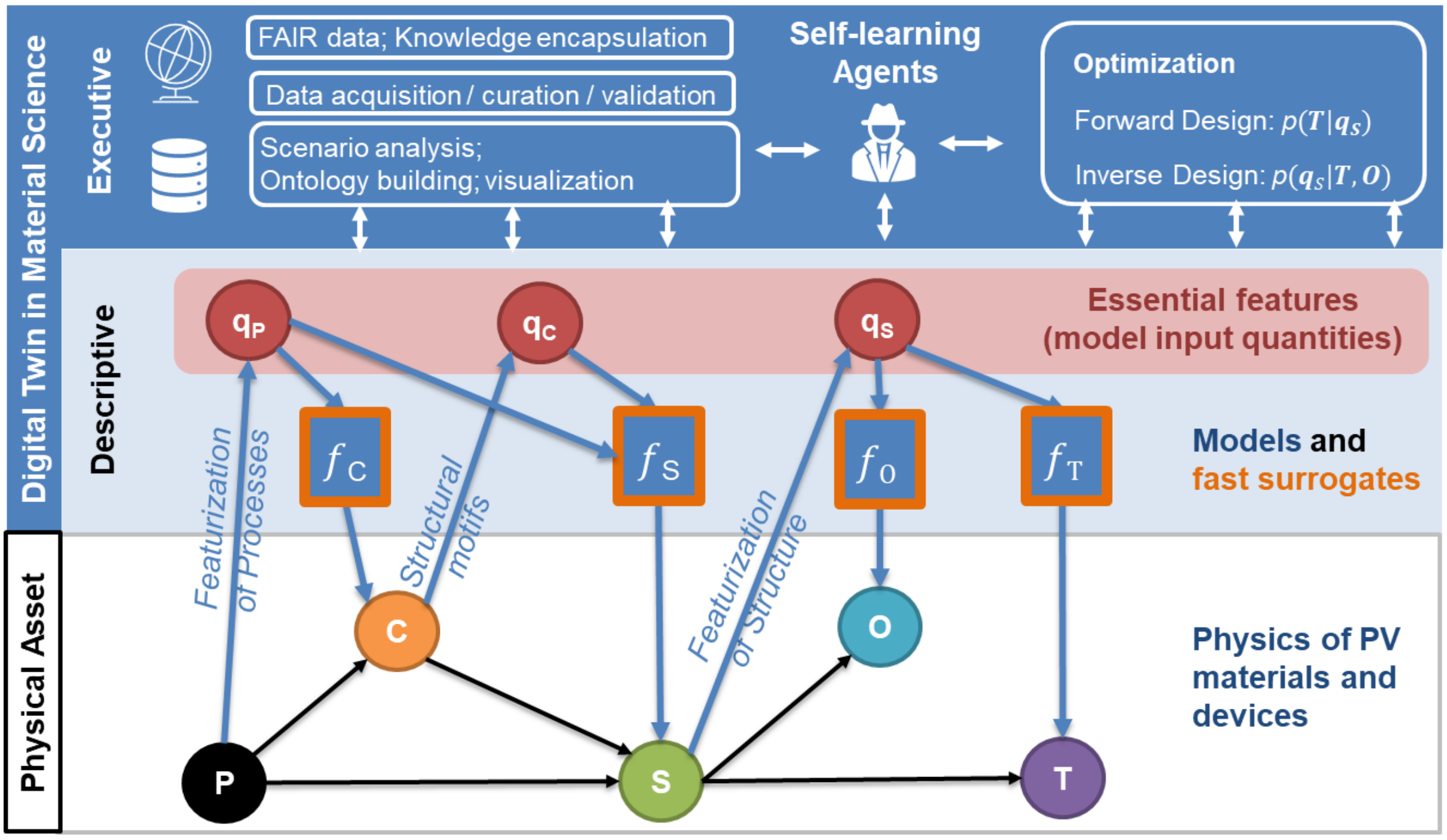

*Figure 4: Definition of a digital twin in material science. White area: physical asset, with causal pathways as described in Figure 3. Light blue area: Physical models and essential features; dark blue area: algorithms that autonomously interact with models and data, improving predictive capacity.*

In Figure 4, we present a layout for an adaptive Digital Twin for photovoltaic materials, derived from the general Digital Twin Structure Model (DTSM,[36]). We define it as an *ensemble of models and solution algorithms autonomously interacting with simulations and experimental interrogations of the physical asset, which represents the full value-added chain of photovoltaics from molecules to PV systems in society, enabling forward and inverse design.* The latter can then be enhanced by optimization approaches for corresponding process design. This layout has the goal to address the challenges to PV material science by a community-driven approach to achieve breakthrough innovations that have the potential to bring novel PV materials into volume markets.

The layout in Figure 4 is aimed at enabling a decisive acceleration over current MAP implementations. First, as mentioned before, the high dimensional problem of predicting function from chemical structure is achieved in two sequential steps of lower dimensionality. Second, a further dimensionality reduction of both **S** and **P** is achieved by encountering the essential features able to “describe the asset well enough to make predictions about its present and future state” [34]. Thus, “featurization of structure” is achieved by an optimized procedure matching experimental data with predictions from fast surrogates that are derived by a detailed and fundamental structural understanding of the underlying physics. And third, an ontology driven approach allows data sharing according to the principles of FAIR data management, allowing massive experimental evidence to build the featurization of structure.

Figure 4 shows the interrelation between the physical asset (lower region with white background) and the Digital Twin (light and dark blue shaded areas). The **descriptive laye**r (light blue region) contains models obtained both from data as well as from physical laws and the parameters they depend upon. As **T** is defined by **S**, and **S** is defined by **P** and **C**, we can learn the featurization of structure independently from chemical structure, by finding the function $\boldsymbol{T} = f_T(\boldsymbol{q_S})$, where $\mathbf{q}_\mathbf{S}$ is a vector of model input quantities (essential structural features) to yield **T**. The essential structural features are first handcrafted (domain anisotropy, tortuosity etc.) and then extended by agents in a dynamic top-down approach, in which first macroscopic features are inferred from comparing the predictive models $p(\mathbf{T}|\mathbf{O})$ **with** macroscopic device simulations (see chapter 4b). When the data basis is solid enough, which is captured by uncertainty quantification (UQ), then mesoscopic and finally microscopic quantities are inferred. At this point, $f_T$ and $\mathbf{q}_\mathbf{S}$ will hold general predictive capacity for **T** from **S**, meaning that whenever we synthesize chemical structures **C** and apply process conditions **P** that yield **S**, we will – within the known uncertainty limits from UQ – observe **T**. Furthermore, Figure 4 shows that the set of essential structural features $\mathbf{q}_\mathbf{S}$ not only determines **T** but also **O**, via the function $\boldsymbol{O} = f_O(\boldsymbol{q_S})$. In chapter 4b, we will exploit this fact by building fast proxy experiments which pave the way to a large and varied dataset to develop fast surrogates to the computational expensive models $f_T$ and $f_O$. The last step could be done by implementing different training strategies, but it could also be based on physical modelling.

Availability of predictive models for **T** from microscopic and mesoscopic structure simplifies the main task of the Digital Twin, namely inverse molecular design, to encounter structural motifs $\mathbf{q}_\mathbf{C}$ that cause microscopic and mesoscopic structure **S** (see chapter 4c). To train the model $\boldsymbol{S} = f_S(\boldsymbol{q_C})$, we will need to calculate and validate microscopic structure from chemical structure, for which fast and reliable methods are not yet available. Based on recent achievements in ML-enhanced modelling, phase field simulation and optimisation, the Digital Twin will allow to advance the existing QM/MM tools to handle disordered materials in the solid state.

Finally, we will achieve a featurization of processes by finding the function $\boldsymbol{C} = f_C(\boldsymbol{q_P})$ able to predict chemical reactions from essential features $\mathbf{q}_\mathbf{P}$ describing process conditions. The resulting predictive models will thus yield *changes* of **C** (due to reactions) with concomitant changes in **S** and therefore **T**, allowing to predict degradation.

In the **executive layer** (dark blue region in Figure 4), autonomous agents enable the Digital Twin to interact with the physical asset and with the community of researchers. Agents are algorithms for specific tasks, autonomously learning under uncertainty to improve performance [38,39], see also [62]. For engineering Digital Twins, building blocks for the executive layer are known and functional [36, 37] and a general model foundation for autonomously attaining predictive capacity has been presented [35]. The algorithms presented in Figure 4, however, must go beyond these known implementations because in PV material science we cannot only rely on existing purely data-based or purely physics-based models. Instead we will need the executive layer to build integrated models for us, by enabling existing solid state models to access relevant scales which in turn requires novel optimization methods to be developed (see chapter 4b).

The tasks of the agents are to autonomously interact with the connected MAP to maximize knowledge gain from new evidence (exploration) and to direct experimentation towards the desired **T** (exploitation). Importantly, out-of-distribution (OOD) samples are autonomously detected by comparison with existing data and decisions are taken whether to disregard the evidence or whether to run dedicated experimental campaigns to encounter new physics. The agents also propel the featurization of structure by optimization of a growing set of essential features against a growing pool of evidence from observations. This is also the point where human experience enters the procedure, using visualization to allow validating newly encountered features, in the sense of building extension models to existing models [36].

The executive layer of the Digital Twin also performs knowledge encapsulation. The hierarchy of microscopic, mesoscopic, and macroscopic structural features (see Figure 3) naturally forms an ontology which can be streamlined by collaboration with national or international initiatives for FAIR ("findable, accessible, interoperable, reusable") data management.

## 4 Empowering a Digital Twin in PV materials

When discussing the experimental results in Figure 2, we found that we can expect PV materials with radically improved properties if we are able to discover molecules and processing conditions that make opposing structural motifs act in concert. In principle, discovery of such molecules can proceed entirely in a black box fashion: generative models and graph neural networks are able to encounter structure-property relationships directly from learning device performance from molecular structure. If successful, the "latent space" of these models will necessarily contain information about the structural motifs that these molecules cause under the given process conditions, because generative models, as any knowledge creation process, are bound to the causal relationships given in Figure 3. However, without prior knowledge about the existence of these opposing structural motifs, the generative models would have to learn them "from scratch", which would involve a huge amount of device data using different molecules of vast structural variation. No research group can provide these data on its own, so the generative models would be confronted with uncertainties about subtle differences in processing methods and ambient conditions in the different labs. These sources of uncertainty, and the sheer amount of device data needed together with the resulting long running time for training, will severely hamper the ability of generative models to attain molecular inverse design capacity.

The acceleration potential of the Digital Twin, in the structure proposed in Figure 3, comes from identifying opposing structural motifs with decisive influence on target properties. This allows the ML methods to be trained directly on the opposing structural motifs, rather than the final target property. Therefore, the ML methods can learn more rapidly to individually optimize opposing requirements, following the idea of known operator learning [45, 48]. In order to identify these opposing structural motifs, the Digital Twin will learn fast proxy experiments to deduce mesoscopic structure from simple to measure optical probes, so that MAPs can create large amounts of varied datasets, reducing the need for formulation of complete devices with concomitant uncertainties. In addition, insights obtained from the ruling physics will significantly increase the quality and secure convergence of the obtained generative models.

### 4a Featurization of structure: identifying decisive structural motifs

Although there are fundamental microscopic (e.g. visible light absorption) and macroscopic (Kirchhoff rules) boundary conditions for PV systems, it is the mesoscopic structure which decides whether opposing requirements are met and thus maximum performance is reached. A prominent example is the formation of nanofibril networks in non-fullerene acceptors yielding PCE values of the corresponding OPV cells close to 20% [30]. Nanofibrils solve the opposing requirements of maximum charge generation (requiring large interfacial area) and maximum charge extraction (requiring small interfacial area) by providing extreme anisotropy of acceptor domains thus orthogonalizing charge generation and extraction. Currently, there is lack of knowledge which molecules tend to form nanofibrils under which conditions, how the performance increase is related to the network shape and its relation to the contacts, and whether these nanofibrils are stable under operational conditions. Investigations into these problems are extremely tedious. On the experimental side, electron microscopy and grazing incidence wide angle X-ray spectroscopy (GIWAXS) are required, techniques which yield exact geometries but at too slow a pace to collect a large and varied dataset of different molecules and process conditions. On the theoretical side, the current parametrization of quantum mechanical and mesoscopic modelling levels of theory is not exact enough to predict the mesoscopic structure formed by a molecule, or if formed, will deteriorate under operational conditions. These considerations hold also for all other kinds of structural motifs, whether known or yet to be discovered.

The Digital Twin, as formulated in Figure 3, can address these challenges on both the experimental and theoretical sides. On the experimental side, it enables the creation of fast proxy experiments on the fly, in order to allow characterization of structural features in a high throughput workflow. Some example workflows are shown in Figure 5. Figure 5a shows the so-called "multi-fidelity" approach, by which a fast proxy experiment is trained by an offline high-fidelity experiment [25]. We can control **P** to obtain a meaningful variation of **S** and then perform a fast but low-fidelity experiment yielding **O,** and a slow but high-fidelity experiment directly yielding the essential features $\mathbf{q}_\mathbf{S}$ that can be used as input parameters to the model $f_T$ which yields **T**, see Figure 4. In this setting, predictive models are obtained by machine learning methods. If experimental data is scarce, data augmentation schemes can be used by generating both **O** and $\mathbf{q}_\mathbf{S}$ using inexpensive low fidelity models, exhaustively sampling the vicinity of the available data, for example by assuming a Gaussian distribution (blue path in Fig. 5a). In a purely data-driven approach, Bash et al. [25] have used graph-based regression to learn to predict electrical conductivity in P3HT-CNT composites from a series of easy to measure probes. Deep learning workflows for multi-fidelity experiments have been successfully deployed in fields beyond disordered semiconductors, for example to extract mechanical material properties from instrumented indentation [41], and to predict the characteristics of human movement from wearable sensors [43]

Another approach to building a fast proxy experiment is parameter extraction by optimizing the outcome from simulations against experimental data, shown in Figure 5b. This approach is chosen if the quantity of interest is experimentally inaccessible at reasonable cost. Model-based parameter inference is prominent in Life Sciences, where intrusive measurements of important body parameters is a burden and can only be done infrequently or not at all, while proxy experiments would allow continuous monitoring. But also in the context of engineering digital twins, parameter inference by matching sensor data to models has been mentioned as method to "estimate what you can't measure" [49]. Model simulations in soft matter (whether body tissue or disordered semiconductors) are often numerical solutions of partial differential equations (PDE). PDE optimization is a large research area where algorithms have been designed and applied with large success, for example in the design of particulate products, see [63]. For restricted settings, analytic solutions can be derived. For example, when ignoring nonlinear and non-local effects in the synthesis of nanoparticles, the process can be described by a linear population balance equation that can be solved analytically. Furthermore, maximizing the obtained yield such that quality guarantees in the particle sizes are met even under uncertain growth rates can be performed by solving an algorithmically tractable convex optimization problem. The price of robustness, i.e., the cost of robust protection against uncertainties, is negligible [61]. As an example in PV technology, efficient sampling methods such as Markov chain Monte Carlo have been combined with emulation of the posterior distribution using Gaussian Processes to keep the amount of PDE evaluations low, while still enabling uncertainty quantification [48].

As another example, fast surrogates for PDEs have been described using Gaussian Processes [51] and Bayesian Physics-Informed Neural Networks (B-PINNs, [52]), see Figure 5c. Methods for physics-informed deep learning solving inverse problems with hidden physics, have been reviewed recently [53]. Furthermore, Graph Network based simulators are proposed to efficiently tackle complex physics [54], and feature selection methods for structure – functionality mapping in PV have been compared [46, 47]. At the current stage, these methods are typically not apt for real-time applications when many scales need to be bridged. Therefore, novel approaches need to be developed both from AI and the physics side.

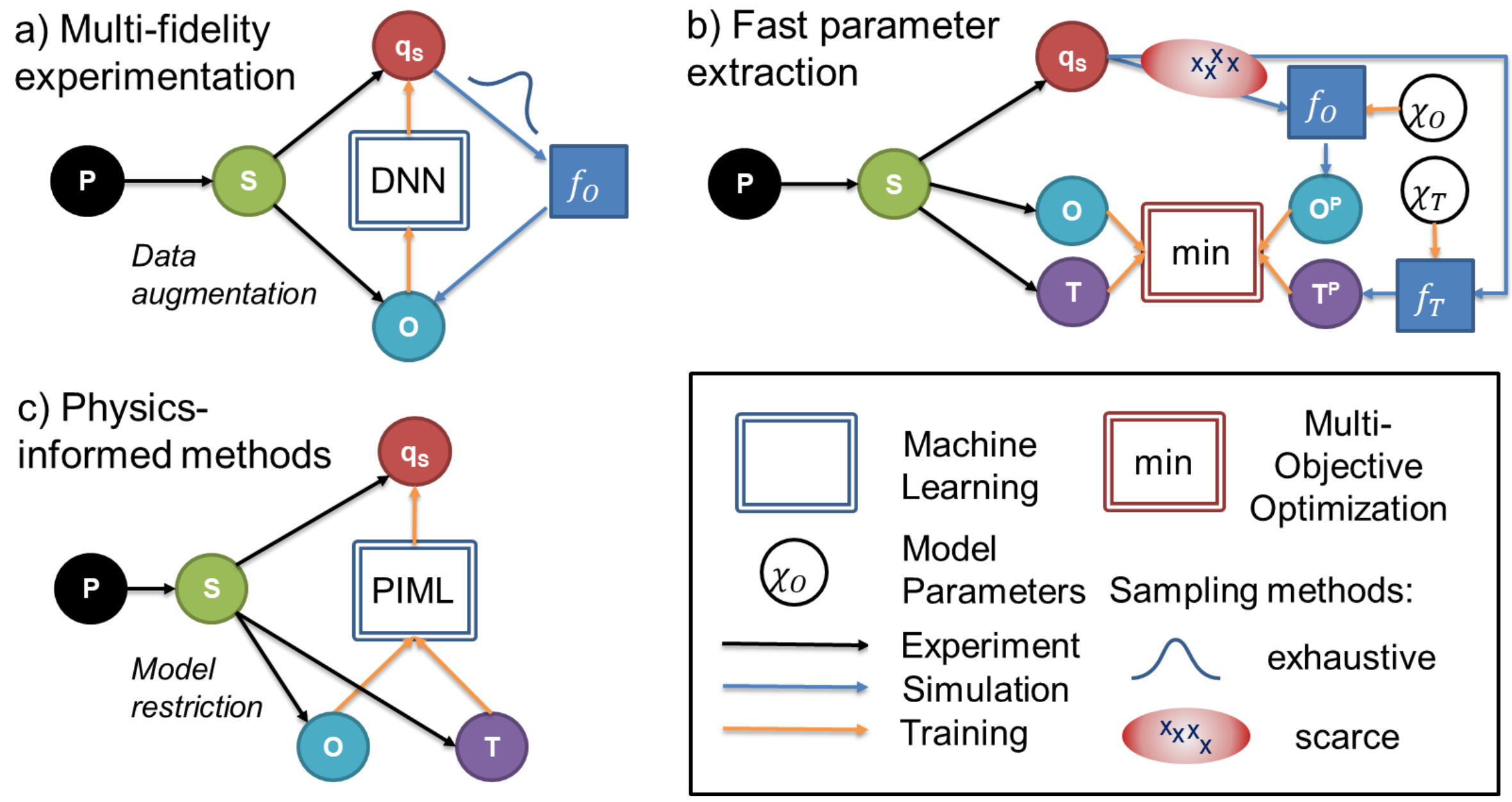

*Figure 5. Encounter of essential features* $\mathbf{q_S}$ *from a) multi-fidelity experimentation under physics constraints, b) fast parameter extraction using multi-objective optimization and efficient sampling for obtaining model parameters* $\chi_{O,F}$ *c) using physics-informed methods.*

## 4b Using big data to improve parametrization of multi-scale models

On the theoretical side, the steady and organized stream of experimental data will allow the Digital Twin to featurize the experiments [55,56] to extract parametrizations of simulations and models obtained from structural insights and physical laws at various scales from the evolution of features. As a vision, this will enable the field of modelling, simulation, and optimization (MSO), to deal with previously inaccessible time and length scales at unprecedented levels of accuracy and generalisation, while addressing uncertainties or unknown information on different scales. In turn, the Digital Twin will benefit from fast and precise forward simulations across scales, greatly alleviating the challenges mentioned in the previous chapter with respect to optimization under uncertainty.

Matter simulations involve typically approximate constitutive equations including Hamilton operators; therefore precise potentials, associated rates or momenta, as appropriate for the underlying model, are of utmost importance. Unfortunately, even ab initio methods require approximations entailing hyperparameters that depend on material classes and thus reduce generalisation, on the other hand, allowing for necessary upscaling. Each restriction in the parameter space, required to accommodate many-body correlations, larger electronic systems or longer time scales, is paid for by further approximations and loss of generality. However, if the models are not general, we cannot use currently existing simulations for automated inverse molecular design, because we need to use it precisely to take us to the unknown where no parametrizations exist. Given the quasi-infinite chemical space, there is no chance to achieve a true generalization of model simulations by whatever means. One possible solution may be to reparametrize the models *as we move*, which is closer to the concept of transfer learning than a true generalization. Such a transfer learning can be very efficiently done by a Digital Twin being driven by self-learning agents deciding on the next experiment or simulation carrying the highest new evidence given prior information.

A particularly relevant example of the use of machine learning to improve modelling comes from DFT, which is a highly active field of research [57]. It has been shown that by adaptive sampling of sparse data regions, the time scale of multiconfigurational photodynamics simulations of a small molecular model system can be extended to 10 ns at high accuracy [40]. Such calculations have the potential to predict photochemical stability, the current Achilles heel of OPV systems. Predicting the excited states in complex environments and morphological stability requires very precise ground state potentials [42] or derived rates [58], which can be parametrized by high-throughput data of D:A systems under varying process conditions via the intermolecular couplings and rate constants the resulting alignments would produce according to DFT calculations, using the methods shown in Fig. 5b. DFT calculations themselves can improve their predictive capacity using high throughput data where little evidence is available today, for example with respect to energies and elementary rate constants at buried interfaces, or with respect to the quenching of triplet states in the solid state, indirectly yielding triplet energies. Together with the fast surrogate replacements of PDE calculations shown in Fig. 5c, we have the perspective for a simulation of a PV system across all length and time scales, with production relevant accuracies and response times.

Optimizations of the kind shown in Figure 5b can be performed as isolated high throughput research projects. In the context of a Digital Twin, they can be executed much more efficiently because a rich and varied dataset will be available for optimization. However, this large variety, accompanied by large variations of veracity, makes learning more difficult because contradicting datasets may wash out patterns that could have been observed in an isolated project. In addition, this knowledge obtained from data needs to be integrated with knowledge from analyzing the underlying physics into the optimization. Addressing the fact that the Digital Twin is prone to measurement errors and insufficient knowledge, it is important to protect the optimized processes and resulting consequences against uncertainties so that reliable results are obtained. including the macroscopic scale into the optimization, binary variables such as the decision between bulk heterojunction or bilayer morphologies, add to the optimization. which is furthermore constrained by physical (solvent orthogonality) or engineering (number of coating stages) constraints. Problems under uncertainty where both binary (or more generally discrete) as well as continuous decisions have to be taken, together with nonlinear functions that represent the underlying physics, belong to the most difficult optimization problems. Typically, they are both difficult in theory and in practice, and solution approaches with quality guarantees have been actively researched in the last decades, see also [64]. Still, challenges consist in the ability to solve large instances quickly. Recently, an optimization approach was presented for discrete-continuous optimization where in addition knowledge learned from data was represented by decision trees. The method has then been applied for the layout of direct current electricity networks [59]. In [62], a data-driven optimization framework was established that can learn the uncertainty together with optimal decisions over time, whenever relevant new data becomes available in an online fashion. Research challenges abound for the layout of the Digital Twin in a tight integration of optimization frameworks with knowledge learned from data over all scales.

For example, they consist in the design of practically efficient algorithms that can optimize under uncertainty or insufficient knowledge, in the ability to integrate new knowledge whenever additional relevant data becomes available, and in the ability of being able to take discrete as well as continuous optimized decisions quickly.

**4c) Chemical structure: generative networks**

Linking chemical structure with microscopic parameters remains the last step to achieve molecular inverse design. We refer to the review by B. Sanchez-Lengeling and A. Aspuru-Guzik [50] showing the achievements that have been made using generative deep learning models. In short, two approaches are prominent, both relying on randomization to efficiently sample the high dimensional search space. In the case of variational autoencoders (VAE), an encoder-decoder architecture is used that generates a latent space in which the essential information (and thus, the physics) is retained to reproduce the encoded entity. To use randomization, chemical structure is represented as probability distribution in the latent space, which allows application of a variational method to find similarity patterns for structures leading to properties. The other approach is generative adversarial networks (GAN), in which a generator produces structure and property from noise, and a detector learns to distinguish fake data from real data. In this way, the latent space is in the generation, and it is getting better the more difficult it gets for the discriminator to tell fake from true. In order to train on experimental properties, the method is combined with reinforcement learning. Challenges consist in the high amount of training data and the difficulties of exploiting and boosting human knowledge because it is difficult to interpret the latent space of the generative models.

In a recent review [60], the potential of graph neural networks (GNN) has been explored for molecular inverse design. It was highlighted that GNN can accommodate environmental effects in the solid state when learning microscopic quantities such as HOMO/LUMO levels from molecular structure; thus, they may be able to address the bottleneck we mentioned in discussing Figure 1b, namely yielding a more efficient transfer of molecule from the microscopic to the mesoscopic stage.

**Conclusions:**

In this work, we have derived the concept of a Digital Twin for photovoltaic materials that can resolve the long-time fundamental challenges in emerging-PV technologies. The Digital Twin concept was split into several building blocks to reduce dimensionality, and causal pathways were distinguished from non-causal relations. The layout is based on existing concepts for digital twins in engineering, and on the current state of the art in materials acceleration platforms (MAP). We motivate the need for a Digital Twin by the need for an increase of the learning curve against an existing and very competitive technique (Si-PV), which however has limits in terms of efficiency and sustainability. The Digital Twin aims at attaining inverse molecular design capacity to discover new materials with unseen properties, matching currently contradicting requirements (interconversion versus deactivation, stability versus processability and recyclability). We propose to approach molecular inverse design in two sequential steps with reduced dimensionality: we first attain predictive capacity for the desired target properties from structure. To this end, the Digital Twin will allow to exploit high throughput capable fast proxy experiments, using cascading surrogates to predict properties across scales. These surrogates integrate models and respective solution approaches obtained from data with those obtained from structural insights taking the underlying physics into account. This will give us access to mesoscopic and microscopic structure, which will reduce the dimensionality of the second problem, finding an ensemble of molecular structures together with optimized process conditions yielding the desired structural properties.

In our opinion, lack of knowledge of mesoscopic and molecular structural parameters is our current bottleneck to the learning rate in PV material science. Our proposed approach is under a big data paradigm. We believe that the needed volume and variability of data cannot be accomplished by a single stakeholder. The ontology-based approach, dictated primarily by the need to accommodate unknown structural and processing features as the data basis evolves, will also enable data sharing following FAIR principles by the entire community, thus creating a vast library of essential structural features with associated process condition. This is the basis for the final goal of achieving full molecular inverse design capacity.

We find that already some of the enabling technologies for our proposed Digital Twin layout are available; however, we point to existing large research gaps. From our point of view, neither the bottom-up modelling nor the data driven approach alone will lead to an efficient Digital Twin with which inverse design and optimized processes for PV material are efficiently possible. Otherwise, either abstract models risk not contain all necessary ingredients, or data-based approaches require too much data or will only show a too limited view on the problem. Instead, we claim that it will always be necessary to truly integrate physics- and data-driven approaches efficiently and along all scales. Then it will be possible to integrate the advantages of both worlds, namely a clear description based on fundamental science for which practically efficient algorithms can be determined, together with data that is able to describe features that may not be contained in a model. Therefore, further research will be necessary for the establishment of approximative chemistry, physics and materials science approaches with data-based integration for surrogates and large-scale mixed integer optimizations, addressing and protecting against uncertainties on all levels. This will allow the Digital Twin to operate at the large scale needed to accomplish its goals.

**Acknowledgements**

A-S S and FL acknowledge support by the Deutsche Forschungsgemeinschaft (DFG, German Research Foundation) for funding within Project-ID 416229255 – SFB 1411 Design of Particulate Products; LL acknowledges financial support by DFG (BR 4031/21-1 and BR 4031/22-1 AOBJ: 681254)

REFERENCES

[1] C. Li, J. Zhou, J. Song, J. Xu, H. Zhang, X. Zhang, J. Guo, L. Zhu, D. Wei, G. Han, J. Min, Y. Zhang, Z. Xie, Y. Yi, H. Yan, F. Gao, F. Liu and Y. Sun, Nature Energy, 2021, 6, 605-613

[2] Q. Liu, Y. Jiang, K. Jin, J. Qin, J. Xu, W. Li, J. Xiong, J. Liu, Z. Xiao, K. Sun, S. Yang, X. Zhang and L. Ding, Science Bulletin, 2020, 65, 272-275

[3] Y. Cui, Y. Xu, H. Yao, P. Bi, L. Hong, J. Zhang, Y. Zu, T. Zhang, J. Qin, J. Ren, Z. Chen, C. He, X. Hao, Z. Wei and J. Hou, Adv Mater, 2021, 33, e2102420

[4] C. J. Brabec, A. Distler, X. Y. Du, H. J. Egelhaaf, J. Hauch, T. Heumueller and N. Li, Advanced Energy Materials, 2020, 10, 2001864

[5] S. Dong, T. Jia, K. Zhang, J. Jing and F. Huang, Joule, 2020, 4, 2004-2016

[6] Z. Jia, Z. Chen, X. Chen, J. Yao, B. Yan, R. Sheng, H. Zhu and Y. Yang, Photonics Research, 2021, 9, 324

[7] Q. Burlingame, M. Ball and Y.-L. Loo, Nature Energy, 2020, 5, 947-949

[8] N. Li, J. D. Perea, T. Kassar, M. Richter, T. Heumueller, G. J. Matt, Y. Hou, N. S. Guldal, H. Chen, S. Chen, S. Langner, M. Berlinghof, T. Unruh and C. J. Brabec, Nature Communications, 2017, 8, 14541

[9] X. Du, T. Heumueller, W. Gruber, A. Classen, T. Unruh, N. Li and C. J. Brabec, Joule, 2019, 3, 215-226

[10] Y. Li, X. Huang, K. Ding, H. K. M. Sheriff, Jr., L. Ye, H. Liu, C. Z. Li, H. Ade and S. R. Forrest, Nature Communications, 2021, 12, 5419

[11] J. Guo, Y. Wu, R. Sun, W. Wang, J. Guo, Q. Wu, X. Tang, C. Sun, Z. Luo, K. Chang, Z. Zhang, J. Yuan, T. Li, W. Tang, E. Zhou, Z. Xiao, L. Ding, Y. Zou, X. Zhan, C. Yang, Z. Li, C. J. Brabec, Y. Li and J. Min, Journal of Materials Chemistry A, 2019, 7, 25088-25101

[12] A. Seemann, T. Sauermann, C. Lungenschmied, O. Armbruster, S. Bauer, H. J. Egelhaaf and J. Hauch, Solar Energy, 2011, 85, 1238-1249

[13] J. Luke, E. M. Speller, A. Wadsworth, M. F. Wyatt, S. Dimitrov, H. K. H. Lee, Z. Li, W. C. Tsoi, I. McCulloch, D. Bagnis, J. R. Durrant and J. S. Kim, Advanced Energy Materials, 2019, 9, 1803755

[14] Y. Wang, J. Lee, X. Hou, C. Labanti, J. Yan, E. Mazzolini, A. Parhar, J. Nelson, J. S. Kim and Z. Li, Advanced Energy Materials, 2020, 11, 2003002

[15] E. M. Speller, A. J. Clarke, J. Luke, H. K. H. Lee, J. R. Durrant, N. Li, T. Wang, H. C. Wong, J.-S. Kim, W. C. Tsoi and Z. Li, Journal of Materials Chemistry A, 2019, 7, 23361-23377

[16] R. Sun, W. Wang, H. Yu, Z. Chen, X. Xia, H. Shen, J. Guo, M. Shi, Y. Zheng, Y. Wu, W. Yang, T. Wang, Q. Wu, Y. Yang, X. Lu, J. Xia, C. J. Brabec, H. Yan, Y. Li and J. Min, Joule, 2021, 5, 1548-1565

[17] X. Xu, J. Xiao, G. Zhang, L. Wei, X. Jiao, H.-L. Yip and Y. Cao, Science Bulletin, 2020, 65, 208-216

[18] Q. Burlingame, X. Huang, X. Liu, C. Jeong, C. Coburn and S. R. Forrest, Nature, 2019, 573, 394-397

[19] https://www.ise.fraunhofer.de/content/dam/ise/de/documents/publications/studies/Photovoltaics-Report.pdf

[20] D.P. Tabor, L.M. Roch, S.K. Saikin, C. Kreisbeck, D. Sheberla, J.H. Montoya, S. Dwaraknath, M. Aykol, C. Ortiz, H. Tribukait, C. Amador-Bedolla, C.J. Brabec, B. Maruyama, K.A. Persson, A. Aspuru-Guzik, Accelerating the discovery of materials for clean energy in the era of smart automation, Nature Reviews Materials. 3 (2018) 5–20. https://doi.org/10.1038/s41578-018-0005-z.

[21] Steven A. Lopez, Benjamin Sanchez-Lengeling, Julio de Goes Soares, Alán Aspuru-Guzik, Design Principles and Top Non-Fullerene Acceptor Candidates for Organic Photovoltaics, Joule 1, 857–870, 2017

[22] Olivier J.J. Ronsin, Jens Harting, Formation of Crystalline Bulk Heterojunctions in Organic Solar Cells: Insights from Phase-Field Simulations, ACS Applied Materials & Interfaces 14 (44), 49785-49800

[23] A Baer, P Malgaretti, M Kaspereit, J Harting, AS Smith, Modelling diffusive transport of particles interacting with slit nanopore walls: The case of fullerenes in toluene filled alumina pores, Journal of Molecular Liquids 368, 120636

[24] Michael C. Heiber, Christoph Baumbach, Vladimir Dyakonov, Carsten Deibel, Encounter-Limited Charge-Carrier Recombination in Phase-Separated Organic Semiconductor Blends, Phys. Rev. Lett. 114, 136602 (2015)

[25] Daniil Bash, Yongqiang Cai, Vijila Chellappan, Swee Liang Wong, Xu Yang, Pawan Kumar, Jin Da Tan, Anas Abutaha, Jayce JW Cheng, Yee-Fun Lim, Siyu Isaac Parker Tian, Zekun Ren, Flore Mekki-Berrada, Wai Kuan Wong, Jiaxun Xie, Jatin Kumar, Saif A. Khan, Qianxiao Li, Tonio Buonassisi, and Kedar Hippalgaonkar, Multi-Fidelity High-Throughput Optimization of Electrical Conductivity in P3HT-CNT Composites, Adv. Funct. Mater. 2021, 31, 2102606

[26] Stefan Langner, Florian Häse, José Darío Perea, Tobias Stubhan, Jens Hauch, Loïc M. Roch, Thomas Heumueller, Alán Aspuru-Guzik, and Christoph J. Brabec, Beyond Ternary OPV: High-Throughput Experimentation and Self-Driving Laboratories Optimize Multicomponent Systems, Adv.Mater.2020, 32, 1907801

[27] Nicola Gasparini, Franco V. A. Camargo, Stefan Frühwald, Tetsuhiko Nagahara, Andrej Classen, Steffen Roland, Andrew Wadsworth7, Vasilis G. Gregoriou, Christos L. Chochos, Dieter Neher, Michael Salvador, Derya Baran, Iain McCulloch, Andreas Görling, Larry Lüer, Giulio Cerullo, Christoph J. Brabec, Adjusting the energy of interfacial states in organic photovoltaics for maximum efficiency, Nature Comms, 12 1772 (2021)

[28] Tobias Ullrich, Dominik Munz and Dirk M. Guldi, Unconventional singlet fission materials, Chem. Soc. Rev., 2021, 50, 3485

[29] Andrej Classen, Lukas Einsiedler, Thomas Heumueller, Arko Graf, Maximilian Brohmann, Felix Berger, Simon Kahmann, Moses Richter, Gebhard J. Matt, Karen Forberich, Jana Zaumseil, and Christoph J. Brabec, Absence of Charge Transfer State Enables Very Low VOC Losses in SWCNT:Fullerene Solar Cells, Adv. Energy Mater. 2019, 9, 1801913

[30] Lei Zhu, Ming Zhang, Jinqiu Xu, Chao Li Jun Yan, Guanqing Zhou, Wenkai Zhong, Tianyu Hao, Jiali Song, Xiaonan Xue, Zichun Zhou, Rui Zeng, Haiming Zhu, Chun-Chao Chen, Roderick C. I. MacKenzie, Yecheng Zou, Jenny Nelson, Yongming Zhang, Yanming Sun,Feng Liu, Single-junction organic solar cells with over 19% efficiency enabled by a refined double-fibril network morphology, Nature Mat. 21 (2022) 656-663

[31] X. Du, L. Lüer, T. Heumueller, J. Wagner, C. Berger, T. Osterrieder, J. Wortmann, S. Langner, U. Vongsaysy, M. Bertrand, N. Li, T. Stubhan, J. Hauch and C. J. Brabec, Joule, 2021, 5, 495-506.

[32] Z. Zhao, R. Anand, M. Wang, IEEE Inter. Conf. Data Sci. Adv. Anal., 2019, 442.

[33 Benjamin Sanchez-Lengeling and Alán Aspuru-Guzik, Inverse molecular design using machine learning: Generative models for matter engineering, Science 361, 360–365 (2018)

[34] a) S. T. Turner, P. Pingel, R. Steyrleuthner, E. J. W. Crossland, S. Ludwigs, D. Neher, Adv. Funct. Mater. 2011, 21, 4640; b) F. C. Spano, J. Chem. Phys. 2005, 122, 234701; c) J. Clark, C. Silva, R. H. Friend, F. C. Spano, Phys. Rev. Lett. 2007, 98, 206406.

[35] Michael G. Kapteyn, Jacob V.R. Pretorius, Karen E. Willcox, A probabilistic graphical model foundation for enabling predictive digital twins at scale, Nature Comp. Sci. 1, 337-347 (2021)

[36] T. Lechler, Jonathan Fuchs, Martin Sjarov, Matthias Brossog, Andreas Selmaier, Florian Faltus, Toni Donhauser, Jörg Franke, Introduction of a comprehensive Structure Model for the Digital Twin

in Manufacturing, 2020, 25th IEEE International Conference on Emerging Technologies and Factory Automation (ETFA), Vienna, Austria, 2020, pp. 1773-1780, doi: 10.1109/ETFA46521.2020.9212030.

[37] Adam Thelen, Xiaoge Zhang, Olga Fink, Yan Lu, Sayan Ghosh, Byeng D.Youn, Michael D. Todd, Sankaran Mahadevan, Chao Hu and Zhen Hu, A Comprehensive Review of Digital Twin - Part 1: Modeling and Twinning Enabling Technologies, arXiv:2208.14197v2 (2022)

[38] Dominik Neumann, Tommaso Mansi, Lucian Itu, Bogdan Georgescu, Elham Kayvanpour, Farbod Sedaghat-Hamedani, Ali Amr, Jan Haas, Hugo Katus, Benjamin Meder, Stefan Steidl, Joachim Hornegger, Dorin Comaniciu, A self-taught artificial agent for multi-physics computational model personalization, Medical Image Analysis 34 (2016) 52–64

[39] Jethro Akroyd, Sebastian Mosbach, Amit Bhave, Markus Kraft, The National Digital Twin of the UK – a knowledge-graph approach, Cambridge University Press, 2020

[40] Jingbai Li, Patrick Reiser, Benjamin R. Boswell, André Eberhard, Noah Z. Burns, Pascal Friederich, Steven A. Lopez Automatic discovery of photoisomerization mechanisms with nanosecond machine learning photodynamics simulations, Chem. Sci., 2021, 12, 5302

[41] Lu Lu, Ming Daob, Punit Kumarc, Upadrasta Ramamurty, George Em Karniadakis, and Subra Suresh, Extraction of mechanical properties of materials through deep learning from instrumented indentation, Proc. Nat. Ac. Sci. 2020, 117(13) 7052-7062

[42] Z. Brkljača, M. Mališ, D. M. Smith, and A.-S. Smith, J. Chem. Theory Comput. 10, 3270-3279, 2014

[43] Eva Dorschky, Marlies Nitschke, Christine F. Martindale, Antonie J. van den Bogert, Anne D. Koelewijn and Bjoern M. Eskofier , CNN-Based Estimation of Sagittal Plane Walking and Running Biomechanics From Measured and Simulated Inertial Sensor Data, Frontiers in Bioengineering and Biotechnology 8 604 (2020); doi: 10.3389/fbioe.2020.00604

[44] Florian Häse, Loïc M. Roch, Christoph Kreisbeck, and Alán Aspuru-Guzik, ACS Central Science 2018 4 (9), 1134-1145, DOI: 10.1021/acscentsci.8b00307

[45] Maier, A.K., Syben, C., Stimpel, B. et al. Learning with known operators reduces maximum error bounds. Nat Mach Intell 1, 373–380 (2019)

[46] a) Pengfei Du, Adrian Zebrowski, Jaroslaw Zola, Baskar Ganapathysubramanian, Olga Wodo, Microstructure design using graphs, npj Computational Materials (2018) 50; b) Hao Liu, Berkay Yucel, Daniel Wheeler , Baskar Ganapathysubramanian, Surya R. Kalidindi, Olga Wodo, How important is microstructural feature selection for data-driven structure-property mapping?, MRS Communications (2022) 12:95–103

[47] Balaji Sesha Sarath Pokuri, Sambuddha Ghosal, Apurva Kokate, Soumik Sarkar and Baskar Ganapathysubramanian, Interpretable deep learning for guided microstructure-property explorations in photovoltaics; npj Computational Materials (2019) 95

[48] A. Maier, H. Köstler, M. Heisig, P. Krauss, S. H. Yang, Known operator learning and hybrid machine learning in medical imaging—a review of the past, the present, and the future, 2022 Prog. Biomed. Eng. 4 022002, DOI: 10.1088/2516-1091/ac5b13

[49] D.R. Gunasegaram, A.B. Murphy, A. Barnard, T. DebRoy, M.J. Matthews, L. Ladani, D. Gu, Towards developing multiscale-multiphysics models and their surrogates for digital twins of metal additive manufacturing, Additive Manufacturing 46 (2021) 102089

[50] L. Mihaela Paun, Dirk Husmeier, Markov chain Monte Carlo with Gaussian processes for fast parameter estimation and uncertainty quantification in a 1D fluid-dynamics model of the pulmonary circulation, Int J Numer Meth Biomed Engng. 2021;37:e3421, https://doi.org/10.1002/cnm.3421

[51] Maziar Raissi and George Em. Karniadakis, Machine Learning of Linear Differential Equations using Gaussian Processes, arXiv:1701.02440v1 (2017)

[52] Liu Yang, Xuhui Menga, George Em Karniadakisa, B-PINNs: Bayesian Physics-Informed Neural Networks for forward and Inverse PDE Problems with Noisy Data, Journal of Computational Physics,Volume 425, 15 January 2021, 109913

[53] George Em Karniadakis, Ioannis G. Kevrekidis, Lu Lu, Paris Perdikaris, Sifan Wang and Liu Yang, Physics-informed machine learning. Nat Rev Phys 3, 422–440 (2021), https://doi.org/10.1038/s42254-021-00314-5

[54] Alvaro Sanchez-Gonzalez, Jonathan Godwin, Tobias Pfaff, Rex Ying, Jure Leskovec, Peter W. Battaglia, Learning to Simulate Complex Physics with Graph Networks, Proceedings of the 37 th International Conference on Machine Learning, Online, PMLR 119, 2020.

[55] F. Hilpert, P.-C. Liao, E. Franz, V. Koch, L. Fromm, E. Topraksal, A. Görling, A.-S. Smith, M. Barr, J. Bachmann, O. Brummel, J. Libuda, ACS Applied Materials & Interfaces, 2023

[56] V. M. Koch, J. Charvot, Y. Cao, C Hartmann, R. G. Wilks, I. Kundrata, I. M-B M-Bacho, N. Gheshlaghi, F. Hoga, T. Stubhan, W. Alex, D. Pokorný, E. Topraksal, A-S Smith, C. J. Brabec, M. Bär, D. M. Guldi, M. K. S. Barr, F. Bureš, and J. Bachmann, Chem. Mater., 34, 21, 9392–9401, 2022

[57] Pascal Friederich, Florian Häse, Jonny Proppe, Alán Aspuru-Guzik, Machine-learned potentials for next-generation matter simulations, Nature Materials 20 (2021) 750-761

[58] T. Bihr, F.-Z. Sadafi, U. Seifert, R. Klupp Taylor, and A.-S. Smith, Adv. Mater. Interfaces 4, 1600310, 2017

[59] D. Gutina, A. Bärmann, G. Roeder, M. Schellenberger, F. Liers Optimization over decision trees: a case study for the design of stable direct-current electricity networks, Optimization and Engineering (2023), DOI 10.1007/s11081-023-09788-x

[60] Patrick Reiser, Marlen Neubert, André Eberhard, Luca Torresi, Chen Zhou, Chen Shao, Houssam Metni, Clint van Hoesel, Henrik Schopmans, Timo Sommer, Pascal Friederich, Graph neural networks for materials science and chemistry, Nature Communications Materials (2022)3:93, https://doi.org/10.1038/s43246-022-00315-6

[61] J. Dienstbier, K. Aigner, J. Rolfes, W. Peukert, D. Segets, L. Pflug, and F. Liers. Robust optimization in nanoparticle technology: A proof of principle by quantum dot growth in a residence time reactor. Comput. Chem. Eng., (2022):157. https://doi.org/10.1016/j.compchemeng.2021.107618.

[62] K. Aigner, A. Bärmann, K. Braun, F. Liers, S. Pokutta, O. Schneider, K. Sharma, S. Tschuppik Data-driven Distributionally Robust Optimization over Time. INFORMS Journal on Optimization (accepted) https://doi.org/10.48550/arXiv.2304.05377

[63] https://www.crc1411.research.fau.eu/

[64] M. Kuchlbauer, F. Liers, M. Stingl. Outer Approximation for Mixed-Integer Nonlinear Robust Optimization, Journal of Optimization Theory and Applications, https://doi.org/10.1007/s10957-022-02114-y (2022).

**Defining and Empowering a Digital Twin in Photovoltaic Materials**

ELECTRONIC SUPPLEMENTARY INFORMATION

Larry Lüer[1], Marius Peters[2], Ana Sunčana Smith[3,4], Eva Dorschky, Bjoern M. Eskofier[5], Frauke Liers[6], Jörg Franke[7], Martin Sjarov[7], Matthias Brossog[7], Dirk Guldi[8], Andreas Maier[9], Christoph Brabec[1,2]

[1] Institute of Materials for Electronics and Energy Technology (i-MEET), Friedrich-Alexander-Universität Erlangen-Nürnberg, Martensstrasse 7, 91058 Erlangen, Germany

[2] High Throughput Methods in Photovoltaics, Forschungszentrum Jülich GmbH, Helmholtz Institute Erlangen-Nürnberg for Renewable Energy (HI ERN), Immerwahrstraße 2, 91058 Erlangen, Germany

[3] PULS Group, Department of Physics, Friedrich-Alexander-Universität Erlangen-Nürnberg, IZNF, Erlangen, Germany.

[4]Division of Physical Chemistry, Ruđer Bošković Institute, Bijenička cesta 54, Zagreb, Croatia.

[5] Machine Learning and Data Analytics Lab, Friedrich-Alexander-Universität Erlangen-Nürnberg, Carl-Thiersch-Straße 2b, 91052 Erlangen, Bayern, Germany

[6] Department of Data Science (DDS), Friedrich-Alexander-Universität Erlangen-Nürnberg, Cauerstr. 11, 91058 Erlangen, Germany

[7] Institute for Factory Automation and Production Systems (FAPS), Friedrich-Alexander-Universität Erlangen-Nürnberg, Egerlandstr. 7, 91058 Erlangen, Germany

[8] Department of Chemistry and Pharmacy, Egerlandstr. 3, Friedrich-Alexander-Universität Erlangen-Nürnberg 91058 Erlangen, Germany

[9] Informatics Department, Friedrich-Alexander-Universität Erlangen-Nürnberg, Martenstr. 3, 91058 Erlangen, Germany

## A Causal relationships for the prediction of $V_{oc}$ from process conditions

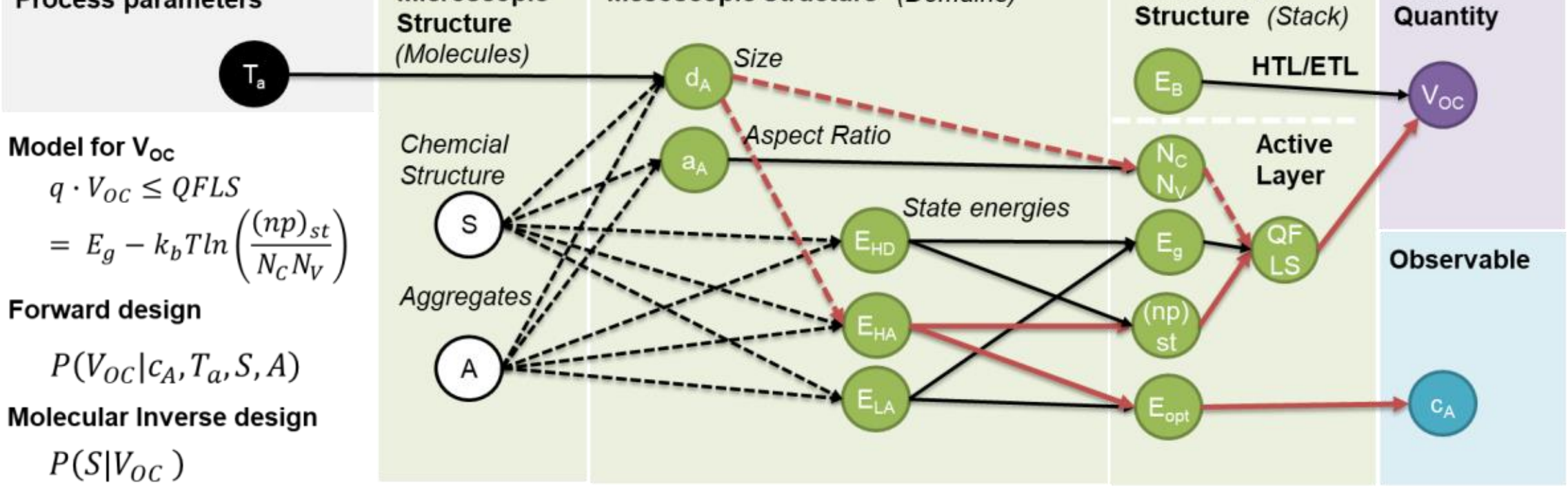

Figure S1: Causal relationships linking process parameters and chemical structure to mesoscopic and macroscopic structure, and macroscopic structure to observables and target quantities, explaining the observed correlations in Figure 2 of main text. The causal relationships are developed from the formula for the quasi-Fermi Level splitting, shown in the left part of the figure. The graph structure shows that inverse design capacity strictly requires knowledge of mesoscopic structure.